\documentclass[10pt]{article} 
\usepackage[preprint]{rlc}

\usepackage{amssymb}            
\usepackage{mathtools}          
\usepackage{mathrsfs}           
\mathtoolsset{showonlyrefs}     
\usepackage{graphicx}           
\usepackage{subcaption}         
\usepackage[space]{grffile}     
\usepackage{url}             
\usepackage{xurl}
\usepackage{rotating}
\usepackage{enumitem}

\usepackage{tabularx, booktabs, multirow}
\newcolumntype{L}{>{\raggedright\arraybackslash}X}

\newtoggle{final}
\togglefalse{final} 

\def\email{\small\ttfamily}
\title{\centering SkillProbe: Security Auditing for Emerging Agent Skill Marketplaces via Multi-Agent Collaboration}

\author{
\parbox{\textwidth}{
\centering
Zihan Guo$^{1,2,\dagger}$, Zhiyu Chen$^{2,3,\dagger}$, Xiaohang Nie$^{2,4}$, Jianghao Lin$^{5}$, 
\\
Yuanjian Zhou$^{2*}$, Weinan Zhang$^{2,5*}$
}
\vspace{0.5em} 
\\
$^{1}$ Sun Yat-sen University \quad$^{2}$ Shanghai Innovation Institute \quad$^{3}$ Tongji University
\\
$^{4}$ Harbin Institute of Technology \quad$^{5}$ Shanghai Jiao Tong University 
\\
$^{\dagger}$ Equal contribution.\quad$^{*}$~Corresponding author.
\\
\email{guozh29@mail2.sysu.edu.cn, wnzhang@sjtu.edu.cn}
}

\begin{document}

\maketitle

\begin{abstract}

With the rapid evolution of Large Language Model (LLM) agent ecosystems, centralized skill marketplaces have emerged as pivotal infrastructure for augmenting agent capabilities. However, these marketplaces face unprecedented security challenges, primarily stemming from semantic-behavioral inconsistency and inter-skill combinatorial risks, where individually benign skills induce malicious behaviors during collaborative invocation. To address these vulnerabilities, we propose SkillProbe, a multi-stage security auditing framework driven by multi-agent collaboration. SkillProbe introduces a ``Skills-for-Skills'' design paradigm, encapsulating auditing processes into standardized skill modules to drive specialized agents through a rigorous pipeline, including admission filtering, semantic-behavioral alignment detection, and combinatorial risk simulation. We conducted a large-scale evaluation using 8 mainstream LLM series across 2,500 real-world skills from ClawHub. Our results reveal a striking popularity-security paradox, where download volume is not a reliable proxy for security quality, as over 90\% of high-popularity skills failed to pass rigorous auditing. Crucially, we discovered that high-risk skills form a single giant connected component within the risk-link dimension, demonstrating that cascaded risks are systemic rather than isolated occurrences. We hope that SkillProbe will inspire researchers to provide a scalable governance infrastructure for constructing a trustworthy Agentic Web.  SkillProbe is accessible for public experience at \url{skillhub.holosai.io}.

\vspace{10pt}
\end{abstract}

\section{Introduction}

The rapid expansion of the Large Language Model (LLM) agent ecosystem has catalyzed a paradigm shift. Systems like OpenClaw\footnote{\url{https://openclaw.ai/}} demonstrate the transformation of agents from simple interfaces into autonomous entities capable of complex, multi-faceted tasks~\citep{jiang2026sokagenticskills}. In this context, a skill acts as an encapsulated unit of procedural knowledge and domain expertise, elevating LLMs from merely invoking isolated APIs to orchestrating semantically-rich, autonomous workflows. Architecturally, a skill consists of documentation, configuration protocols, executable scripts, and required auxiliary assets~\citep{anthropic2025introducingagentskills}.

The architecture decouples semantic decision-making from execution, where agents parse natural language descriptions to decide actions, then trigger low-level scripts. While this semantic-execution decoupling enhances readability and discoverability, it introduces a precarious ``trust propagation chain'', i.e., user intent drives reasoning, descriptions guide invocation directives, and the implementation layer operates on external systems. This trust assumption fails in open marketplaces like ClawHub\footnote{\url{https://clawhub.ai}}, where malicious developers can publish semantically compliant but logically harmful skills~\citep{liu2026maliciousagentskills, jiang2026sokagenticskills}. As the Agentic Web scales, these risks will magnify exponentially, becoming a critical bottleneck for system robustness.

Current defense mechanisms follow two paradigms. The first, agent-centric runtime governance, focuses on real-time intervention. Current methods employ prompt-level defenses such as spotlighting~\citep{hines2024defendingindirectprompt}, while benchmarks quantify agent susceptibility to embedded attacks~\citep{zhan2024injecagentbenchmarkingindirect, cartagena2026mindgaptext}. However, these runtime approaches often fail to neutralize stealthy exploits hidden deep within a skill's logic. The second paradigm, skill-centric static analysis, involves large-scale vulnerability characterization. While existing skill auditing work characterizes vulnerabilities at ecosystem scale~\citep{liu2026agentskillswild}, it remains scoped to discrete code-level patterns and supply chain integrity. Consequently, they are ill-equipped for composite threats that uniquely bridge the semantic and execution layers of the agent ecosystem.

\begin{figure}[tb]
    \centering
    \includegraphics[width=\textwidth]{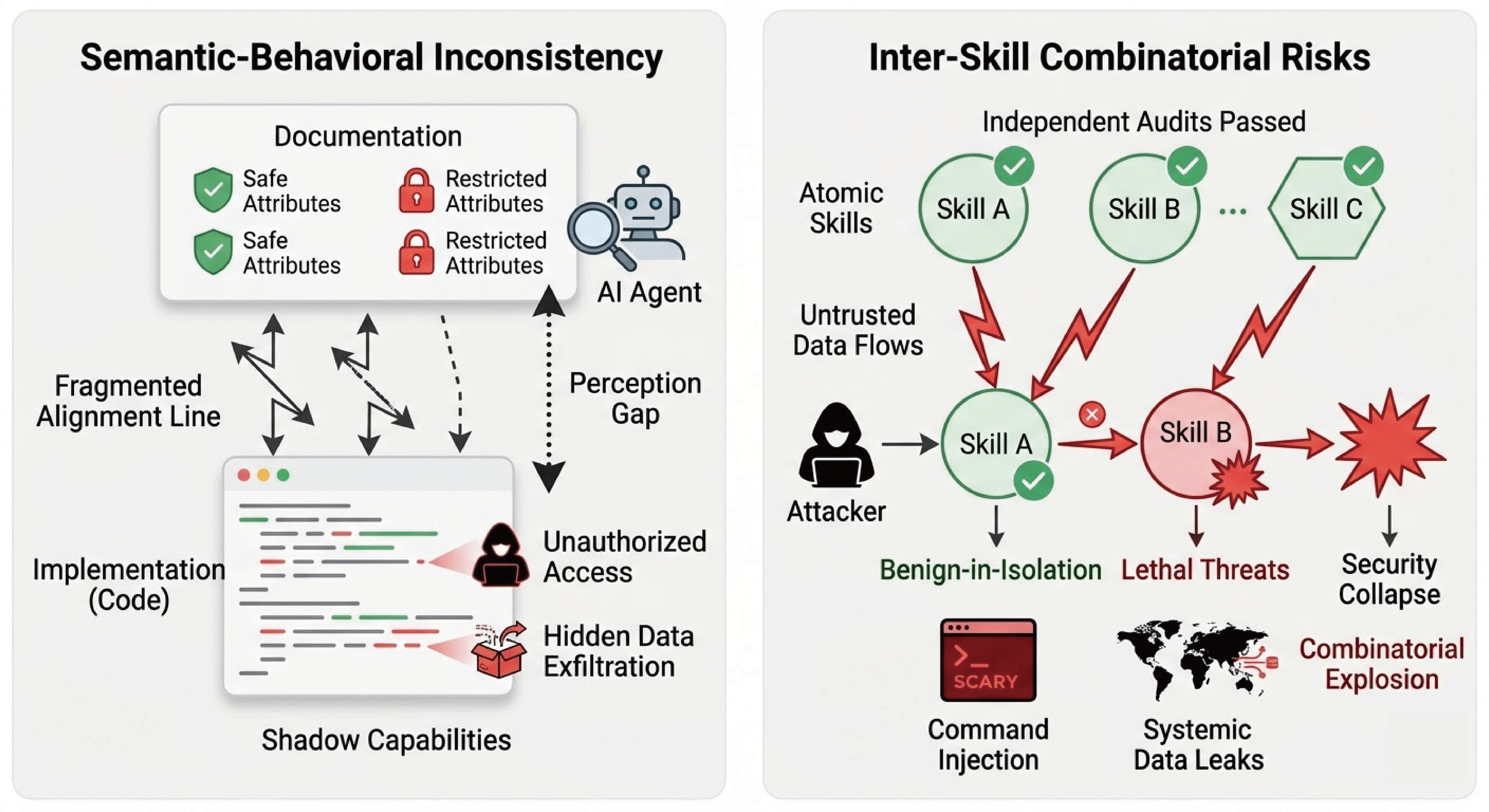}
    \caption{Visualizing invisible risks. Left panel depicts the semantic misalignment between high-level documentation and low-level implementation. While the documentation declares ``safe'' and ``restricted'' attributes, the underlying code harbors shadow capabilities such as unauthorized access and hidden data exfiltration. The fragmented alignment line represents the perception gap caused by the agents' reliance on natural language for black-box invocations. Right panel demonstrates the combinational explosion of risks during cross-skill collaboration. Although individual skills may pass independent audits, their interconnection via untrusted data flows can trigger a security collapse. This effect shows how benign-in-isolation components can mutate into lethal threats, such as command injection or systemic data leaks, when executed in a composite chain.}
    \label{fig:challenge}
\end{figure}

While current methodologies exhibit efficacy in detecting explicit vulnerabilities, they often falter when confronted with the black-box nature of agent decision-making. Since agents rely heavily on natural language descriptions to retrieve and activate skills, which often involves intricate resource interactions and multi-step sequences, two formidable challenges emerge, as shown in Figure~\ref{fig:challenge}:

\begin{enumerate}[nosep,leftmargin=*]
    \item Semantic-Behavioral Inconsistency: A profound gap persists between a skill's documentation and its underlying executable logic. This discrepancy facilitates a novel class of stealthy threats, where implementations may harbor shadow functions or exercise over-privilege beyond their documented scope~\citep{li2026dontbelieveeverything, schmotz2026skillinjectmeasuringagent}. Because an agent's reasoning is anchored in the semantic layer, any execution-layer deviation remains virtually imperceptible to both the user and the agent.
    \item Inter-Skill Combinatorial Risks: Modern agent architectures emphasize the collaborative or parallel invocation of multiple skills. In these dynamic scenarios, skills that appear benign in isolation may induce emergent collaborative attacks when integrated into specific execution chains~\citep{li2026stacwheninnocent, liu2025makeagentdefeat}. Adversaries can engineer skill pairs to leverage misleading outputs from an upstream skill to trigger high-risk behaviors in a downstream counterpart. This risk of dynamic correlation, exacerbated by combinatorial explosion, renders traditional atomic-level auditing ineffective.
\end{enumerate}

To integrate existing security utilities and address the aforementioned emerging challenges, we present SkillProbe, a multi-phase and multi-agent security auditing framework specifically engineered for agent skills. SkillProbe adopts the ``Skills-for-Skills'' design paradigm, wherein each auditing phase is encapsulated as a standardized skill module. This allows general agents to dynamically load and execute these modules on demand. By constructing a three-phase pipeline comprising admission filtering, semantic consistency detection, and combinatorial risk analysis, SkillProbe achieves a synergetic evaluation of both static features and dynamic behaviors of skills. Our key contributions are summarized as follows:

\begin{enumerate}[nosep,leftmargin=*]
  \item Automated Auditing Framework: We introduce SkillProbe, a multi-agent collaborative framework designed for the security auditing of agent skills. It automates the transition from initial compliance filtering and deep semantic consistency analysis to the detection of complex cross-skill link attacks. The framework features a plug-and-play design, enabling the seamless integration of modular third-party security tools and customized rule sets.
  \item Multi-dimensional Risk Modeling: To address semantic inconsistency, we first extract standardized capability features from heterogeneous representations to construct a four-class alignment matrix. Building upon this foundational alignment, we project these capabilities into a standardized label graph to mitigate the challenges of link identification and combinatorial explosion. Finally, by applying risk link policies to this graph, we enable the precise identification of hazardous skill chains emerging from dynamic inter-skill collaborations.
  \item System Implementation and Evaluation: We implemented the SkillProbe prototype using a robust stack that includes FastAPI, Vue 3, and an npm-based REPL (Read-Eval-Print Loop). An extensive longitudinal audit within the ClawHub ecosystem across multiple mainstream LLMs validates the efficacy. Notably, SkillProbe successfully discovered several zero-day vulnerabilities and complex combinatorial attacks that traditional atomic-level auditing failed to detect.
\end{enumerate}

The remainder of this paper is organized as follows: Section 2 establishes the research context for the agent skill ecosystem and provides a systematic review of related work. Section 3 delineates the system design of SkillProbe, detailing the operational mechanisms, core logic, and implementation of its pivotal auditing phases. Section 4 presents the experimental settings and offers a comprehensive analysis of the evaluation results. Section 5 discussed the broader implications of our empirical findings and addressed the limitations of the current framework. Finally, Section 6 concludes the paper and outline directions for future research.

\section{Related Work}

\subsection{Agent Skill Ecosystems and Emerging Threats}

Agent skills, standardized as an open specification by Anthropic~\citep{anthropic2025introducingagentskills}, are self-contained packages comprising a \texttt{SKILL.md} specification, executable scripts, and related assets, loaded via a three-level progressive disclosure architecture that injects only the required context tier into the agent's window.
Community marketplaces such as ClawHub have since enabled open third-party skill publication, establishing the threat environment that motivates this research~\citep{jiang2026sokagenticskills}.

The rapid scale-up of skill marketplaces has introduced a correspondingly serious security challenge.
The first empirical study at the ecosystem-scale~\citep{liu2026agentskillswild} audited 31,132 skills and found that 26.1\% contained at least one vulnerability spanning 14 patterns, with skills combining executable scripts 2.12$\times$ more likely to be vulnerable.
A follow-up ground-truth analysis~\citep{liu2026maliciousagentskills} further confirmed two attacker archetypes (\emph{Data Thieves} and \emph{Agent Hijackers}) operating at scale through behavioral verification of 98,380 skills.
ClawHub, the primary open registry for Anthropic-standard skills, has deployed a post-publication scanning pipeline combining VirusTotal signature matching with an LLM-assisted code analyzer, detecting known malicious payloads, hardcoded credentials, and explicit prompt injection patterns in individual skill bundles~\citep{liu2026maliciousagentskills}. However, this scanning stack operates reactively at the code-artifact level and shares a fundamental blind spot: it cannot verify whether a skill's natural-language documentation faithfully represents its actual executable behavior, nor model emergent risks that arise only when multiple skills are composed into an execution chain.
Critically, both studies explicitly scope out \emph{skill-chaining} scenarios and operate post-hoc on existing marketplaces, leaving two gaps unaddressed, i.e., detecting the semantic-behavioral inconsistency and identifying cross-skill compositional attack chains before registration.

\subsection{Security Threats in LLM Agent Ecosystems}

The proliferation of tool-augmented LLM agents has surfaced a family of ``description-driven'' threats that traditional vulnerability scanners cannot detect.
Indirect prompt injection ~\citep{greshake2023notwhatyouve} embeds adversarial directives inside external data retrieved by agents, allowing attackers to redirect agent behavior without touching the model weights.
Systematic evaluations confirm the severity of this vector. For example, InjecAgent~\citep{zhan2024injecagentbenchmarkingindirect} benchmarks indirect prompt injection across 1,054 test cases spanning 17 user tools and 62 attacker tools, showing how injected instructions in tool responses can redirect agent behavior toward attacker-controlled actions, while the GAP benchmark~\citep{cartagena2026mindgaptext} reveals a systematic divergence between text-level refusals and tool-call compliance across six frontier models, with conditional GAP rates reaching up to 79.3\% under tool-encouraging prompts.
Skill-Inject~\citep{schmotz2026skillinjectmeasuringagent} further demonstrates that skill files constitute a potent injection surface, achieving up to 80\% attack success rate and inducing data exfiltration and ransomware-like behaviors on frontier models.
Runtime defenses, such as spotlighting~\citep{hines2024defendingindirectprompt}, output-level filtering, and defense benchmarking~\citep{yi2025benchmarkingdefendingindirect}, offer partial mitigation, but all operate \emph{after} a malicious invocation has been dispatched, inheriting an irreducible TOCTOU (Time-of-Check to Time-of-Use) window.

More recently, research has shifted toward the supply side of the tool ecosystem.
Empirical studies on MCP-integrated systems find that tool descriptor fields routinely misrepresent underlying behavior, exploiting LLMs' reliance on natural-language schema fields as trusted context~\citep{li2026dontbelieveeverything}.
Together, these findings establish that the description layer (not the executable code) is the primary attack surface in open skill marketplaces.

\subsection{Code--Document Semantic Consistency and API Contract Verification}

Ensuring that a tool's implementation faithfully reflects its documentation has a long history in software engineering.
Early neural approaches detect just-in-time comment inconsistencies at the method level~\citep{panthaplackel2020deepjustintimeinconsistency}.
More recent work extends this to repository-scale checks via LLM-based categorization and filtering~\citep{xu2025docprismlocalcategorization} and fine-tuned code LLMs~\citep{rong2025codecommentinconsistency}.
In the API contract mining space, DAInfer~\citep{wang2024dainferinferringapi} infers API aliasing specifications from library documentation via neurosymbolic optimization, DocFlow~\citep{tileria2024docflowextractingtaint} extracts taint specifications from natural-language software documentation, and RESTSpecIT~\citep{decrop2025youcanrest} automates RESTful API documentation inference and black-box testing via LLM-assisted request mutation.
Despite their breadth, all these methods share a common assumption: the documentation is authored in good faith and the goal is to keep the code consistent with it.

This research operates under the inverse threat model, i.e., the documentation itself may be the attack vector.
A malicious developer can craft a skill whose description appears benign to both the LLM and a human reviewer while the underlying code exfiltrates data or escalates privileges, which is what we term \emph{semantic-behavioral inconsistency} in the above section.
Concurrently, \citep{li2026dontbelieveeverything} measure description--code inconsistency across 10,240 MCP servers, finding $\approx$13\% harbor undocumented high-risk operations.
Their framework, MCPDiFF, however, operates at the implementation-semantics layer and explicitly excludes over-claimed capabilities from its threat model.

\subsection{Compositional Security and Multi-Step Attack Chains}

Atomic auditing of individual tools is insufficient when skills are composed into chains.
Sequential Tool Attack Chains (STAC)~\citep{li2026stacwheninnocent} demonstrate that individually harmless tools can be orchestrated to achieve attack success rates exceeding 90\%.
Taint analysis of LLM agents shows that malicious prompt payloads can traverse multiple tool-call boundaries via the agent's reasoning chain, evading any single-tool policy~\citep{liu2025makeagentdefeat}.
The TOCTOU problem further compounds this, where a skill passing a point-in-time audit may silently diverge at runtime.
Existing agent-security benchmarks such as AgentDojo~\citep{debenedetti2024agentdojodynamicenvironment} provide valuable red-team evaluations of compositional attacks, but they operate in post-deployment settings and do not offer marketplace-level gatekeeping.

\subsection{Multi-Agent Systems for Security Auditing}

Multi-Agent Systems (MAS) have demonstrated clear advantages over monolithic tools in security tasks that benefit from role specialization and parallel task decomposition.
AutoGen~\citep{wu2023autogenenablingnextgen}, MetaGPT~\citep{hong2024metagptmetaprogramming}, and LangGraph~\citep{langchainailanggraphbuild} establish the orchestration primitives, including structured message passing, delegated tool invocation, and stateful workflow graphs, on which complex auditing pipelines can be built.
Collaborative LLM agents have demonstrated the ability to systematically probe attack surfaces at a scale and depth that manual review cannot match.
However, applying MAS to \emph{skill marketplace auditing} introduces a trust boundary problem, where the auditing agents must reason about tools they cannot fully trust, and the orchestration framework itself becomes a target.

\section{System Design}

\begin{figure}[tb]
    \centering
    \includegraphics[width=\textwidth]{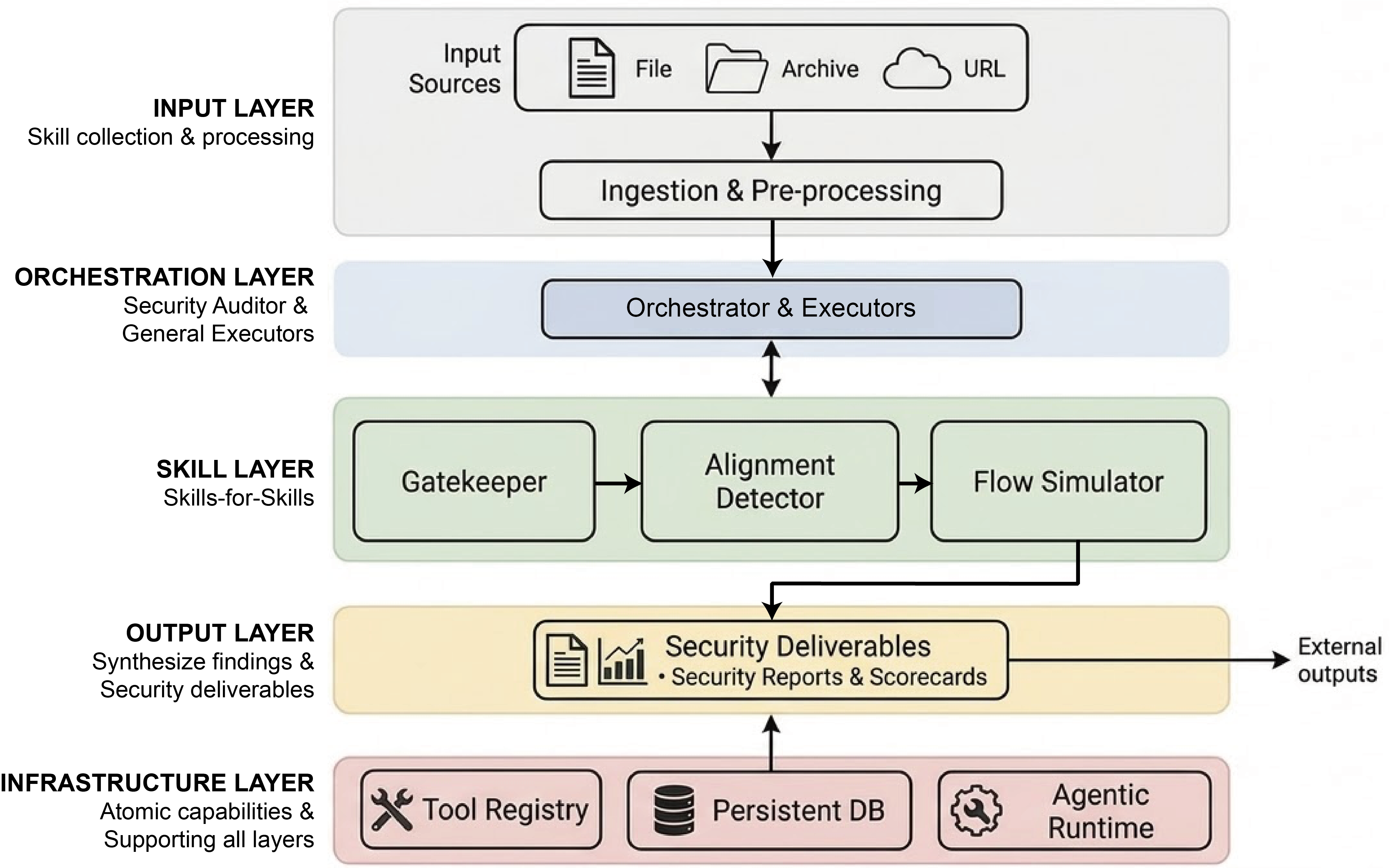}
    \caption{The overall system architecture of SkillProbe. The framework consists of five primary components, i.e., the Input Layer for data ingestion, the Orchestration Layer for multi-agent collaboration and task planning, the Skill Layer for core audit phases, the Output Layer for delivering the auditing findings, and the Infrastructure Layer for underlying storage and computing.}
    \label{fig:architecture}
\end{figure}

\subsection{System Overview}

\subsubsection{SkillProbe Framework}

SkillProbe adopts a hierarchical and modular multi-agent security auditing framework, realizing the self-consistent objective of auditing skills via agents with skills. The framework is organized into five functional layers as illustrated in Figure~\ref{fig:architecture}:

\begin{enumerate}[nosep,leftmargin=*]
    \item Input Layer: Handles the ingestion and pre-processing of auditing targets. It maintains compatibility with heterogeneous physical formats, including local directories, compressed archives, and remote URLs.
    \item Orchestration Layer: Acts as the central dispatching hub. A specialized Security Auditor serves as the lead orchestrator, maintaining the auditing state, distributing tasks, and driving low-level executors through sequential phase-wise scheduling.
    \item Skill Layer: The core functional domain based on the ``Skills-for-Skills'' paradigm. This layer integrates three primary auditing modules, i.e., Gatekeeper for admission filtering, Alignment Detector for semantic consistency, and Flow Simulator for combinatorial risk analysis, along with their respective sub-workflows.
    \item Output Layer: Synthesizes auditing findings into structured security reports and generates a multi-dimensional security scorecard.
    \item Infrastructure Layer: Provides foundational atomic capabilities, including a tool registry, a persistent database for audit fingerprints, and a dedicated agentic loop runtime environment.
\end{enumerate}

\subsubsection{``Skills-for-Skills'' Paradigm}

SkillProbe abstracts each auditing phase into a standardized skill package, embodying the philosophy of ``Specification-as-Code''. Each package encapsulates natural language specifications (defining auditing intent), machine-readable workflow configurations (providing programmatic implementation), system prompts, and operational scripts. Under this design, the auditing process is represented as an agentic collaborative flow based on a Directed Acyclic Graph (DAG). This decoupling allows developers to extend or refine auditing logic simply by updating skill packages without modifying the framework’s core engine, significantly enhancing system extensibility.

\subsubsection{Execution Engine}

The dynamic execution in SkillProbe is orchestrated by a Security Auditor and driven by several general agents, which supports three adaptive modes, including 1) sequential processing of standard serial execution and serial computed batch-calls for efficiency, 2) dynamic concurrency comprising parallel dynamic for real-time scaling and parallel pair for comparative tasks, and 3) pre-calculated parallelism for high-throughput multi-processing. By decoupling execution topology from task logic, the agents ensure high code re-usability and interface uniformity through universal atomic capabilities and on-demand tool loading.

\subsubsection{Auditing Mode}

To balance throughput and analysis depth, SkillProbe offers two operational modes. The first one is quick mode, which executes the Gatekeeper and Alignment Detector serially. It focuses on the rapid identification of salient malicious features and is ideal for high-frequency preliminary screening. Another one is the standard mode, which employs full parallelization across all phases to maximize throughput and ensure comprehensive coverage, suitable for deep and system-level security assessments.

\subsection{Core Auditing Phases}

\begin{figure}[tb]
    \centering
    \includegraphics[width=\textwidth]{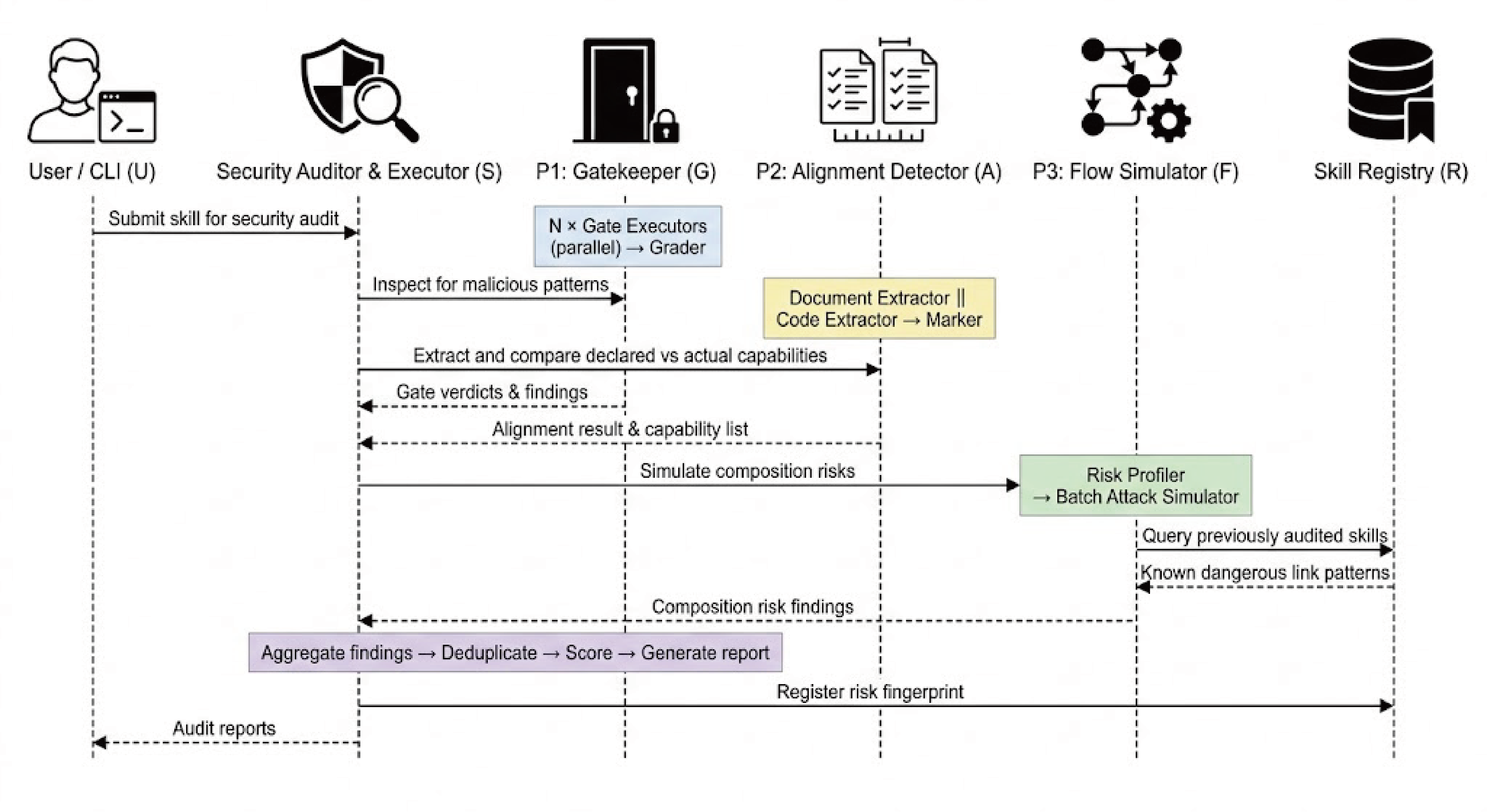}
    \caption{The overall workflow of the SkillProbe security auditing framework. The process comprises three main phases: (P1) malicious pattern inspection via the Gatekeeper, (P2) capability alignment detection to compare declared vs. actual functions, and (P3) composition risk simulation within the Flow Simulator. The Security Auditor coordinates these tasks, eventually generating a comprehensive report and registering the skill's risk fingerprint in the registry.}
    \label{fig:workflow}
\end{figure}

The workflow of SkillProbe is illustrated in Figure~\ref{fig:workflow}, which can be divided into three auditing phases.

\subsubsection{Phase 1: Gatekeeper}

In the skill ingestion phase, the system evaluates multiple dimensions, including compliance, malicious code patterns, hazardous dependencies (e.g., packages with known CVEs), and the rationality of permission declarations. Recognizing that a single agent may suffer from cognitive lacunae when processing heterogeneous evaluation logic, Gatekeeper employs the architecture of decoupled inspection and centralized aggregation.

Existing security utilities are encapsulated as autonomous Gate Executors for parallel scrutiny. For instance, the integrated Skill Vetter\footnote{\url{https://github.com/openclaw/skills/tree/main/skills/spclaudehome/skill-vetter}} scans for backdoors and permission escalations. To ensure zero-omission of critical risks, deterministic detection mechanisms are integrated alongside LLM-based analysis. Upon completion, a Grader agent pre-parses the outputs from all Gate Executors into structured dictionaries, injecting them into the context for zero-latency aggregation. The aggregation follows a conservative principle, i.e., a single ``BLOCK'' decision from any gate results in a global ``BLOCK'' status, while a skill proceeds to the next phase only if it achieves a ``PASS'' across all dimensions.

\subsubsection{Phase 2: Alignment Detector}

Consistency issues are formalized as follows. Let $D(s)$ be the set of capabilities declared in the documentation of skill $s$, and $C(s)$ be the set of capabilities implemented in its code. Each capability $c$ is represented as a tuple $c = \langle \text{Resource}, \text{Access} \rangle$, denoting the resource type and its corresponding access level. The objective of the Alignment Detector is to precisely categorize the intersection and union relationships between $D(s)$ and $C(s)$, highlighting high-risk deviations.

The auditing process involves parallel execution by a Document Extractor and a Code Extractor. The former utilizes regex-guided extraction to derive an explicit capability list from natural language, while the latter employs pluggable code analysis tools combined with AST analysis and LLM reasoning to perform I/O Binding Analysis, tracing the dependencies between sensitive call parameters and return values. Subsequently, a Marker agent applies a semantic normalization algorithm to map these capabilities to a canonical key. We define a four-class alignment matrix to categorize risk states:

\begin{enumerate}[nosep,leftmargin=*]
    \item Match ($D = C$): Full alignment.
    \item Over-declaration ($D \supset C$): Excessive permissions claimed.
    \item Under-declaration ($C \supset D$): Presence of shadow functions.
    \item Mixed ($C \cap D \neq \emptyset, C \setminus D \neq \emptyset, D \setminus C \neq \emptyset$): Complex deviations.
\end{enumerate}

\subsubsection{Phase 3: Flow Simulator}

To address risks emerging from collaborative skill invocation, the Flow Simulator employs risk fingerprint tagging. This approach significantly reduces the search complexity from an exponential $O(2^N)$ brute-force enumeration to a linear $O(N \times |\text{Rules}|)$ rule-matching process, enabling scalable risk detection.

Specifically, a Risk Profiler agent infers risk labels employing a strategy that prioritizes deterministic rules while utilizing semantic inference as a supplement. It assigns output risk tags (characterizing the output features of a skill) and input sensitivity tags (describing downstream sink risks). To converge the search space, the system implements a dual pre-filtering mechanism consisting of resource type validation and structural constraints. For example, document-only skills are filtered for code execution tag. Based on these tags, the system executes batched attack path simulations guided by risk link policies, covering command injection, indirect prompt injection, and data exfiltration. To ensure fidelity, we introduce an evidence enforcement mechanism that requires agents to cite specific snippets from the original code or documentation for every identified vulnerability.

\subsection{Auditing Verdict}

To provide an intuitive and granular representation of auditing results, SkillProbe introduces a security scorecard to evaluate a target skill across three pivotal dimensions:

\begin{enumerate}[nosep,leftmargin=*]
    \item Malicious Patterns: Directly derived from the Gatekeeper (Phase 1) findings, this dimension scrutinizes active malicious behaviors such as backdoors, data exfiltration, and dependencies with known CVEs.
    \item Semantic Consistency: Grounded in the Alignment Detector (Phase 2), this metric quantifies the potential attack surface by identifying shadow functions and over-declarations between documentation and executable logic.
    \item Composition Safety: Leveraging the Flow Simulator (Phase 3), this dimension probes for combinatorial risks and simulates attack chains emerging from inter-skill collaborations.
\end{enumerate}

The detection status of each dimension is determined by the severity of the identified findings. The final verdict follows a strict veto policy (one-vote veto), categorizing skills into three levels:

\begin{enumerate}[nosep,leftmargin=*]
    \item REJECTED: Indicates the presence of critical vulnerabilities or severe semantic conflicts.
    \item CONDITIONAL: Denotes low-risk discrepancies or findings that necessitate manual review.
    \item APPROVED: Assigned when all metrics strictly adhere to the established security baselines.
\end{enumerate}

\subsection{System Implementation}

To support large-scale deployment and practical utility, we developed the Holos-SkillHub platform\footnote{\url{skillhub.holosai.io}}, comprising a high-performance backend, a reactive web interface, a Command-Line Interface (CLI), and integration plugins for LLMs.

The core system is powered by a FastAPI-based backend and a Vue 3 frontend, delivering a closed-loop auditing service from asynchronous task scheduling to multi-dimensional security visualization. The backend utilizes the asyncio framework for efficient task queue management. We integrated embedding models to generate 1,536-dimensional semantic vector indexes, complemented by a SHA256 file fingerprinting mechanism to ensure global de-duplication and rapid retrieval of auditing targets. Built with Naive UI, the frontend provides an intuitive dashboard. It employs the WebSocket protocol for real-time telemetry of auditing progress. Complex analyses are rendered via auditing certificates and inter-skill risk topology maps, transforming raw simulation data into actionable decision-support insights.

To bridge the gap between development and governance, SkillProbe provides two specialized tools.  One is the Command Line Interface (CLI) named Holos-SkillHub. It's a productivity-oriented TypeScript terminal tool that manages the full lifecycle of skill management—from authentication and batch submission to the streaming download of multi-format reports. Another one is a OpenClaw plugin. This plugin introduces the ``Conversation-as-Auditing'' paradigm. It allows users to asynchronously trigger auditing workflows and retrieve structured results within LLM environments via lightweight script sequences. This integration significantly lowers the barrier to entry for agentic security governance without requiring complex host environment configurations.

\section{Evaluation}

\subsection{Experimental Setup}

To rigorously validate the efficacy and robustness of SkillProbe, we constructed an evaluation environment grounded in a real-world ecosystem.

The empirical dataset was curated from the ClawHub platform, comprising a large-scale collection of agent skills as of March 16, 2026. To ensure the representativeness and practical impact of our samples, we performed sampling based on a descending order of download volume. This strategy prioritizes skills with higher prevalence and potential influence within the ecosystem.

To evaluate the auditing consistency across diverse reasoning capacities and code comprehension levels, we integrated 8 predominant LLM series as the underlying reasoning engines:

\begin{enumerate}[nosep,leftmargin=*]
    \item High-Reasoning/Coding-Specialized Models: Claude Sonnet 4.6\footnote{\url{https://www.anthropic.com/claude/sonnet}}, Gemini 3.1 Pro\footnote{\url{https://deepmind.google/models/gemini/pro/}}, GPT‑5.2‑Codex\footnote{\url{https://developers.openai.com/api/docs/models/gpt-5.2-codex}}, and Nex-N1.1\footnote{\url{https://huggingface.co/nex-agi/DeepSeek-V3.1-Nex-N1.1}}.
    \item Efficient/Lite Models: Claude Haiku 4.5\footnote{\url{https://www.anthropic.com/news/claude-haiku-4-5}}, Gemini 3.1 Flash-Lite\footnote{\url{https://deepmind.google/models/model-cards/gemini-3-1-flash-lite/}}, GPT-5 mini\footnote{\url{https://developers.openai.com/api/docs/models/gpt-5-mini}}, and Nex-N1\footnote{\url{https://huggingface.co/nex-agi/DeepSeek-V3.1-Nex-N1}}.
\end{enumerate}

\begin{table}[t]
\centering
\caption{Core risk link policies for combinatorial attack detection.}
\label{tab:risk-link-policy}
\small
\renewcommand{\arraystretch}{1.1}
\begin{tabularx}{\columnwidth}{c L L L}
\toprule
\textbf{ID} & \textbf{Output Risk (Source)} & \textbf{Input Sensitivity (Sink)} & \textbf{Attack Pattern} \\
\midrule
P1 & Unsanitized external string & Shell/Interpreter execution & Command Injection \\
P2 & Unsanitized external string & LLM prompt injection & Indirect Prompt Injection \\
P3 & Unsanitized external string & Sensitive action parameter & Parameter Tampering \\
\midrule
P4 & Executable code fragments & LLM prompt injection & Indirect Prompt Injection \\
P5 & Executable code fragments & Shell/Interpreter execution & Remote Code Execution \\
\midrule
P6 & Sensitive credentials/data & External network egress & Data Exfiltration \\
P7 & Sensitive credentials/data & LLM prompt injection & Data Leakage \\
\midrule
P8 & Manipulative semantics & Persistent state/config & Fact Poisoning \\
P9 & Manipulative semantics & Sensitive action trigger & Intent Hijacking \\
\bottomrule
\end{tabularx}
\end{table}

The Gatekeeper employs the Skill Vetter policy as the baseline for admission filtering. All experiments were conducted using SkillProbe’s standard mode to maintain an optimal balance between analysis depth and computational efficiency. 
And to address the complexities of inter-skill collaborative risks, we established a set of core risk link policies, summarized in Table \ref{tab:risk-link-policy}. This policy dictates the pairing rules between output risk tags (source) and input sensitivity tags (sink). Our model encompasses 9 typical attack paradigms, spanning from traditional software vulnerabilities to agent-specific threats such as indirect prompt injection and intent hijacking.

\subsection{Cross-model Audit Applicability}

To evaluate the generalization and consistency of SkillProbe across diverse reasoning engines, we curated a benchmark comprising the top 20 most downloaded skills. This dataset spans core categories such as information retrieval, coding assistance, and workflow automation. Paired with 8 state-of-the-art models, this resulted in a total of 160 auditing samples.

\begin{table}[htbp]
  \centering
\caption{Per-model audit statistics across ClawHub top 20 skills.}
  \label{tab:model-comparison}
  \resizebox{\textwidth}{!}{%
  \begin{tabular}{lrrrrrr}
    \toprule
    \textbf{Model} & \textbf{Avg. Time (s)} & \textbf{Approved (\%)} & \textbf{Conditional (\%)} & \textbf{Warn. Findings} & \textbf{Info Findings} & \textbf{Avg. Recs} \\
    \midrule
    Claude Haiku 4.5 & 86.7 & 80\% & 20\% & 4 & 41 & 0.5 \\
    Claude Sonnet 4.6 & 228.6 & 55\% & 45\% & 10 & 82 & 1.7 \\
    Gemini 3.1 Flash-Lite& 18.5 & 90\% & 10\% & 2 & 34 & 0.6 \\
    Gemini 3.1 Pro & 221.8 & 80\% & 20\% & 8 & 44 & 1.4 \\
    GPT-5 mini & 171.3 & 85\% & 15\% & 9 & 88 & 1.9 \\
    GPT-5.2-Codex & 120.3 & 95\% & 5\% & 2 & 77 & 2.3 \\
    Nex-N1 & 38.7 & 90\% & 10\% & 2 & 33 & 0.7 \\
    Nex-N1.1 & 282.0 & 95\% & 5\% & 1 & 35 & 0.9 \\
    \bottomrule
  \end{tabular}%
  }  
  \begin{flushleft}
  \footnotesize\textit{
  Avg. Time: mean elapsed seconds for completed (non-blocking) audits. \\
  ~Approved/Conditional: share of 20 skills by final scorecard verdict. \\
  ~Warn./Info: total findings by severity across 20 skills. \\
  ~Avg. Recs: mean number of actionable recommendations per skill.
  }
  \end{flushleft}
\end{table}

Table \ref{tab:model-comparison} summarizes the core metrics of each model. The results reveal significant heterogeneity in performance across the model spectrum. Execution times exhibit cross-order-of-magnitude variances. Gemini-Flash (18.5s) and Nex-N1 (38.7s) demonstrate superior computational efficiency, whereas Nex-N1.1 (282.0s) and Sonnet-4.6 (228.6s) with long-time thinking incur the highest latency, with the latter showing significant sensitivity when parsing high-complexity scripts. Regarding verdict bias, Sonnet-4.6 yielded a ``CONDITIONAL'' rate of 45\%, substantially higher than other models. reflecting fundamental divergences in security thresholds and semantic deviation perception among models. 

Moreover, a distinct auditing trilemma emerges, where achieving high stringency (e.g., Sonnet-4.6), high granularity in findings (e.g., GPT-5-mini), and low latency (e.g., Gemini-Flash) simultaneously is exceedingly challenging. For instance, while GPT-5.2-Codex adopts a more permissive stance, it leads in the average number of remediation recommendations, indicating a preference for providing non-blocking optimization guidance over outright rejection. In contrast, Nex-N1 achieves an optimal equilibrium between response velocity and coverage.

\begin{figure}[tb]
    \centering
    \includegraphics[width=\textwidth]{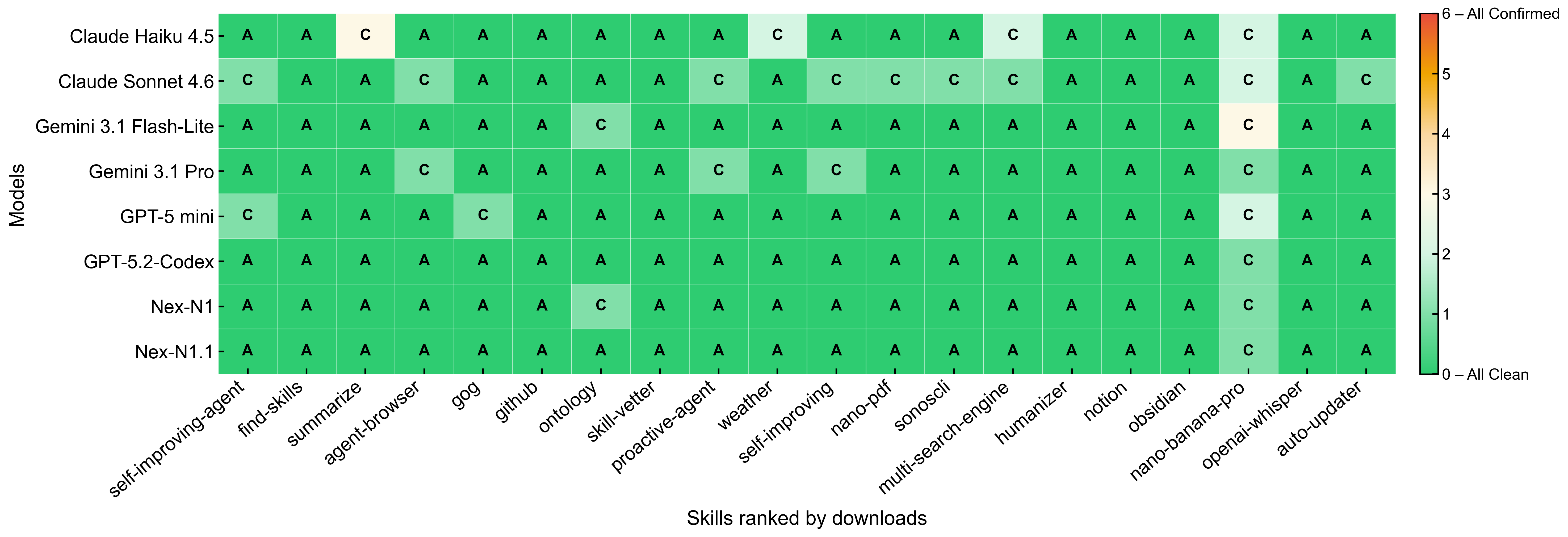}
    \caption{Cross-model audit consistency heatmap for ClawHub top 20 Skills. Each cell represents the aggregated risk score for a pair of model and skill, computed as the sum of per-dimension severity scores (\textit{malicious patterns}, \textit{semantic consistency}, \textit{composition safety}), where \textsc{Clean}$=0$, \textsc{Suspected}$=1$, and \textsc{Confirmed}$=2$ (range: 0--6). Cell labels A and C denote the final scorecard verdict (\textsc{Approved} and \textsc{Conditional}, respectively). Skills are ordered by download count (left to right, descending).}
    \label{fig:consistency-heatmap}
\end{figure}

Figure \ref{fig:consistency-heatmap} illustrates the cross-model audit verdict heatmap. Each cell’s color intensity is determined by a weighted score derived from malicious patterns. For the majority of high-profile skills (e.g., find-skills, github), the columns exhibit a uniform green distribution with consistent A verdicts. This indicates a robust convergent consensus among models regarding the safety of high-quality skills. A notable outlier is nano-banana-pro, which was consistently flagged as C across all eight models. This underscores the presence of systemic semantic deficiencies reproducible across different reasoning engines, demonstrating SkillProbe’s resilience in identifying core risks. Localized color fluctuations in entries like agent-browser and self-improving highlight the vulnerability of edge cases to specific model logical predispositions. 

As for model-wise tendency, Sonnet-4.6 exhibits the highest frequency of C verdicts (9/20), confirming its role as the ``strictest auditor'', particularly adept at unearthing subtle vulnerabilities in semantic alignment. Conversely, the row distributions for Gemini-Flash and Nex-N1.1 are the most monochromatic, suggesting a more lenient auditing style. In summary, the heatmap reveals that while systemic biases in perceptual sensitivity exist between models, SkillProbe maintains high cross-model robustness in identifying high-risk skills. This effectively mitigates the risk of false negatives or positives stemming from the limitations of any single model.

\subsection{Large-scale Empirical Audit}

To evaluate the performance of SkillProbe in a live production environment, we conducted a deep scan of the top 2,500 most-downloaded skills. We selected nex-n1 as the primary auditing engine, leveraging its superior long-context window and high-concurrency reasoning capabilities. This represents the vanguard of the current Agentic Web ecosystem, encompassing both official selections and high-frequency community modules.

\begin{figure}[tb]
    \centering
    \includegraphics[width=\textwidth]{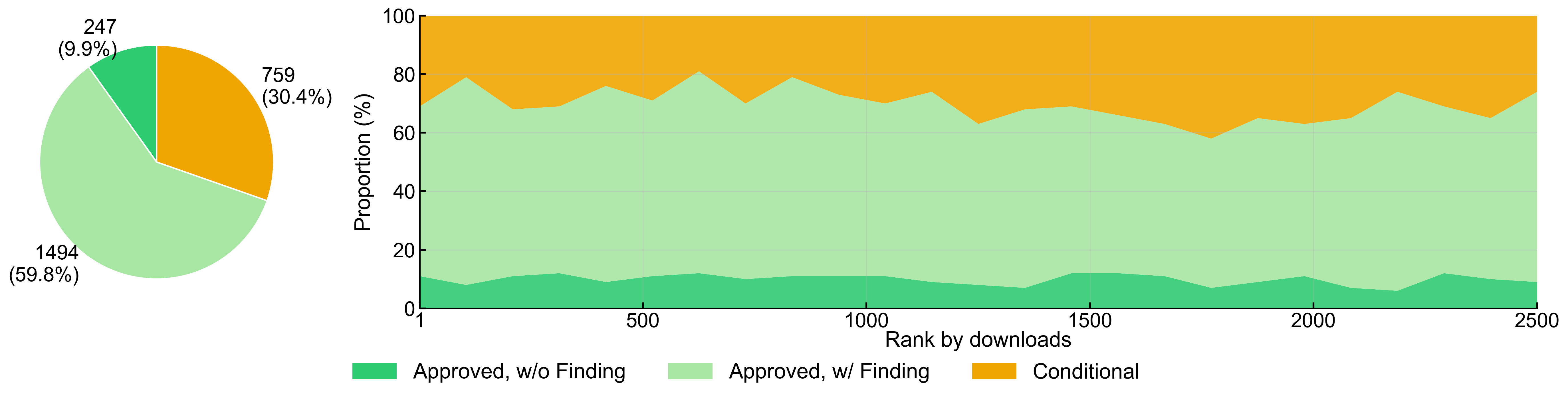}
    \caption{Security audit overview of the top 2500 most-downloaded skills on ClawHub. Left panel shows the distribution of final audit verdicts across 2,500 skills, including 9.9\% are fully clean (Approved, w/o Finding), 59.8\% pass with minor findings (Approved, w/ Finding), and 30.4\% have potential risks (Conditional). Right panel depicts the stacked area chart showing the proportion of each verdict category across 25 equal-sized bins sorted by download rank (1 = most downloaded). The composition remains broadly stable across all popularity tiers, indicating that download rank is not a reliable proxy for security quality.}
    \label{fig:overview-plot}
\end{figure}

As illustrated in Figure~\ref{fig:overview-plot}, we performed a quantitative analysis of the auditing results across two dimensions, namely overall proportion and rank distribution. Our findings reveal a sobering reality of overall security stance. Among the 2,500 leading skills, only 247 (9.9\%) are in a fully ``clean'' state (i.e., approved without any findings). While 59.8\% passed the baseline audit, their reports still contain observable minor security flaws. Most critically, 30.4\% (759 skills) were marked as ``Conditional'' due to violations of security rules. This indicates that over 90\% of popular skills failed to meet rigorous security admission standards, which is a high risk-ratio that imposes extreme demands on the boundary defense.

Note that the stacked area chart unveils a counter-intuitive phenomenon, where the proportion of conditional skills exhibits a striking statistical invariance across different download tiers, consistently hovering between 20\% and 35\%. Despite the fact that high-download skills undergo more frequent community scrutiny and iterative updates, their risk profiles do not significantly improve as their ranking rises. This finding decisively refutes the intuitive assumption that popularity equals security, proving that automated auditing is equally indispensable for all ecological tiers.

\begin{figure}[tb]
    \centering
    \includegraphics[width=\textwidth]{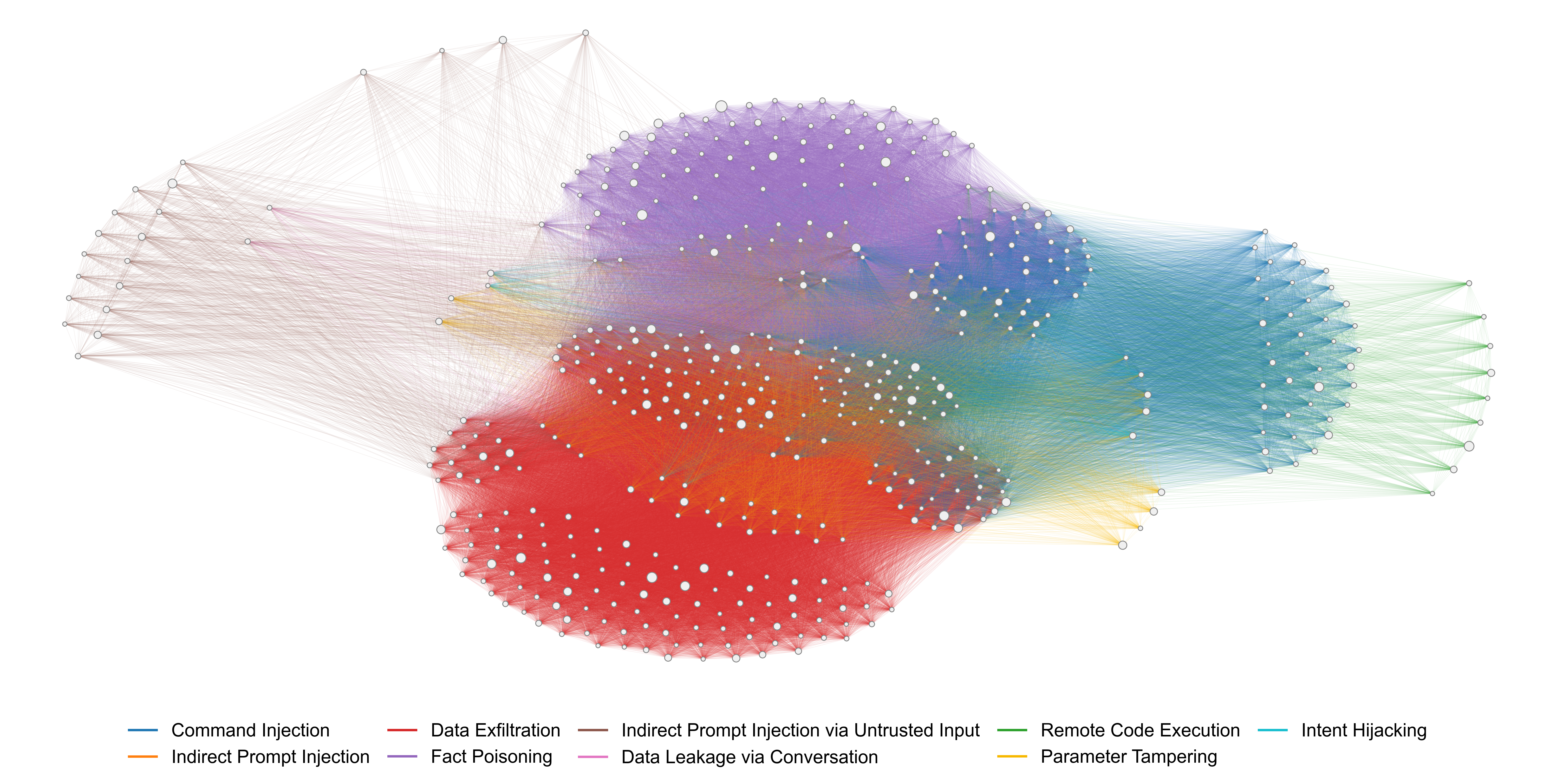}
    \caption{Risk chain network of the 499 skills with at least one confirmed exploit from the nine official attack chains. Each node represents one skill, with node size scaling with download count. Edge color encodes the shared risk chain type between two connected skills (nine colors corresponding to nine risk categories as illustrated in Table \ref{tab:risk-link-policy}). Edges connect any two skills that share at least one confirmed chain type.}
    \label{fig:risk-chain}
\end{figure}

For the 499 skills confirmed to contain at least one officially defined risk link, we constructed a risk-association network, as shown in Figure \ref{fig:risk-chain}. The network comprises 75,373 risk edges, exhibiting high centrality. Data Exfiltration (highlighted in red, 32,328 edges) and Fact Poisoning (highlighted in purple, 28,441 edges) dominate the network, accounting for 80\% of total edges. This suggests that the majority of high-risk skill pairs are driven by sensitive data leakage paths and untrusted content persistence paths. Command injection (blue, 8,637 edges) ranks third, primarily concentrated in functional modules that execute system instructions. The uniform distribution of node sizes further confirms that high-risk features are not confined to niche skills but are pervasive across high-popularity modules.

The most critical discovery is the structural connectivity. The entire high-risk subset is fully interconnected in the risk dimension. From a defensive standpoint, this fully connected structure signifies a multiplicative effect on the attack surface rather than a simple linear addition. Once any two high-risk skills are combined in a complex task, they can easily trigger cross-link attacks via shared hazardous capability patterns. This empirical result provides irrefutable support for the necessity of SkillProbe's cross-skill combinatorial simulation phase, i.e., traditional atomic auditing tools are fundamentally blind to such cascaded risks within a fully interconnected threat landscape.

\section{Discussion}

\paragraph{The Popularity Paradox and the Imperative for Market Standardization} One of the most striking empirical revelations of this paper is the significant decoupling between a skill's popularity (download volume) and its actual security posture. Our results indicate that even within the top 100 elite skills, the prevalence of potential risks remains on par with long-tail modules. This popularity paradox exposes a structural deficit in the security governance of current agent marketplaces (e.g., ClawHub). Users often conflate high download counts with high quality, while developers, under the pressure of rapid iteration, frequently bypass rigorous security scrutiny. This finding underscores that in a burgeoning Agentic Web ecosystem, an automated and multi-dimensional admission auditing protocol like SkillProbe is not a mere luxury but a foundational infrastructure for a trustworthy ecosystem. Future marketplaces should mandate a standardized multi-dimensional certification system to eliminate the ``blind trust'' engendered by popularity metrics.

\paragraph{The Semantic Gap beyond Code-Level Scrutiny} While traditional security auditing predominantly targets logical vulnerabilities within the code, this research demonstrates that semantic inconsistency has emerged as a critical threat vector in the agent skill ecosystem. Since an agent’s selection of a skill is heavily predicated on its natural language description while execution is anchored in script logic, this cross-modal discrepancy provides a natural stealth path for malicious actors. Even if the code itself lacks conventional backdoors, an adversary can achieve malicious objectives by crafting descriptions that induce the agent into unintended invocations. The findings from SkillProbe urge the security community to shift focus from isolated vulnerability scanning to intent-behavior alignment verification, elevating semantic auditing to a status of parity with traditional code auditing.

\paragraph{Paradigm Shift from Atomic Inspection to Combinatorial Defense} The giant connected component revealed in our risk-link map highlights the risk of cascading failures within the agent ecosystem. When high-risk skills achieve such pervasive interconnectivity, the vulnerability of a single module can exert multiplicative destructive power through the skill chain. Traditional atomic auditing is fundamentally ill-equipped to capture dynamic threats that only manifest during skill composition. The existence of over 75,000 risk edges suggests that a complex ``risk highway'' has already formed within the ecosystem. Consequently, future defensive architectures must evolve from inspecting individual plugins to simulating entire execution chains. The Flow Simulator proposed in this work validates the feasibility of combinatorial risk prediction under linear complexity, providing a technical template for mitigating the threat of combinatorial explosion.

\paragraph{Limitations and Threat Modeling} Despite the robust auditing capabilities of SkillProbe, several limitations warrant further investigation.

\begin{enumerate}[nosep,leftmargin=*]
    \item Detection Depth: For highly obfuscated code or entirely black-box third-party API responses, the precision of the Code Extractor remains constrained by the reasoning depth of the underlying LLM~\citep{li2025everything}.
    \item Zero-day Patterns: Current combinatorial simulations rely on predefined risk link policies. Hence, a detection blind spot may exist for novel, undefined zero-day attack patterns~\citep{Moamin_Abdulhameed_Al2025}.
    \item Computational Overhead: The balance between auditing stringency and latency remains a challenge when facing a massive instantaneous influx of skills (e.g., >10,000)~\citep{park2025survey}.
\end{enumerate}

\section{Conclusion}

This paper presents SkillProbe, a collaborative multi-agent framework designed to bridge the critical gaps in the security governance of the agent skill ecosystem. By formalizing the semantic gap and addressing the formidable challenges of risk combinatorial explosion, SkillProbe shifts the security paradigm from reactive runtime interventions to proactive, intent-aware pre-distribution auditing. The framework’s ``Skills-for-Skills'' design ensures exceptional extensibility and modularity, facilitating the seamless integration of evolving third-party security utilities.

Our large-scale empirical evaluation based on ClawHub provides the first systematic evidence of the popularity paradox, revealing that high-download skills harbor latent risks comparable to those of long-tail modules. Furthermore, the discovery of a single giant connected risk component underscores the multiplicative effect of threats within multi-skill orchestration environments, exposing the inherent limitations of traditional atomic-level auditing in perceiving cascaded risks.

SkillProbe establishes a robust foundation for constructing a secure and trustworthy Agentic Web. Future research will focus on incorporating dynamic sandboxing techniques to deeply scrutinize obfuscated code logic~\citep{triedman2025multi} and leveraging reinforcement learning to adaptively discover zero-day combinatorial attack patterns~\citep{naeem2025cyber}. As agent autonomy continues to advance, automated collaborative auditing frameworks like SkillProbe will become indispensable components in safeguarding Agent-driven workflows.

\bibliography{main}
\bibliographystyle{rlc}

\end{document}